\documentclass[screen,sigplan,twocolumn]{acmart}
\acmSubmissionID{261}
\renewcommand\footnotetextcopyrightpermission[1]{}
\AtBeginDocument{%
  }
\AtBeginDocument{\pagestyle{plain}}

\setcopyright{none}

\usepackage{xspace}
\usepackage{amsmath}
\usepackage{subcaption}
\usepackage[font=small]{caption}
\usepackage[shortlabels]{enumitem}
\usepackage{filecontents}
\usepackage{titlesec}
\usepackage{array}
\usepackage[utf8]{inputenc}
\usepackage{pifont}
\usepackage{textcomp}
\usepackage{tabularx}
\newcommand{\difftxt}[1]{#1}

\newcommand{\sysname}{WALI\xspace}
\newcommand{\zephyrsysname}{WAZI\xspace}
\newcommand{\sysnamelower}{wali}
\newcommand{\sysnamebold}{\textbf{WALI}\xspace}

\newcommand{\singlewp}{\textit{1-to-1}\xspace}
\newcommand{\multiwp}{\textit{N-to-1}\xspace}
\newcommand{\threadlesswp}{\textit{threadless}\xspace}
\definecolor{ForestGreen}{HTML}{00994D}

\newcommand{\greencheck}{\textcolor{ForestGreen}{\ding{51}}}
\newcommand{\redx}{\textcolor{red}{\ding{55}}}

\newcounter{qacounter}
\newcommand{\questionanswer}[2] {\refstepcounter{qacounter}
  \paragraph{Q: #1}
  #2
}

\titlespacing*{\paragraph}{0em}{*1}{*1}
\setlist[itemize]{leftmargin=2em,itemsep=0.2em}
\setlist[enumerate]{leftmargin=2em,noitemsep}

\newcolumntype{P}[1]{>{\centering\arraybackslash}p{#1}}

\title[Empowering WebAssembly with Thin Kernel Interfaces]{Empowering WebAssembly with \\ Thin Kernel Interfaces}

\author{Arjun Ramesh}
\orcid{0009-0007-7390-0295}
\affiliation{%
  \institution{Carnegie Mellon University}
  \city{Pittsburgh}
  \state{Pennsylvania}
  \country{USA}
}
\email{arjunr2@andrew.cmu.edu}

\author{Tianshu Huang}
\orcid{0009-0005-6788-1302}
\affiliation{%
  \institution{Carnegie Mellon University}
  \city{Pittsburgh}
  \state{Pennsylvania}
  \country{USA}
}
\email{tianshu2@andrew.cmu.edu}

\author{Ben L. Titzer}
\orcid{0000-0002-9690-2089}
\affiliation{%
  \institution{Carnegie Mellon University}
  \city{Pittsburgh}
  \state{Pennsylvania}
  \country{USA}
}
\email{btitzer@andrew.cmu.edu}

\author{Anthony Rowe}
\orcid{0000-0003-2332-9450}
\affiliation{%
  \institution{Carnegie Mellon University, Bosch Research}
  \city{Pittsburgh}
  \state{Pennsylvania}
  \country{USA}
}
\email{agr@andrew.cmu.edu}

\renewcommand{\shortauthors}{Arjun Ramesh, Tianshu Huang, Ben L. Titzer, Anthony Rowe}

\begin{document}

\begin{abstract}
Wasm is gaining popularity outside the Web as a well-specified low-level binary format with ISA portability, low memory footprint and polyglot targetability, enabling efficient in-process sandboxing of untrusted code.
Despite these advantages, Wasm adoption for new domains is often hindered by the lack of many standard system interfaces which precludes reusability of existing software and slows ecosystem growth.

This paper proposes \emph{thin kernel interfaces} for Wasm, which directly expose OS userspace syscalls without breaking intra-process sandboxing, enabling a new class of virtualization with Wasm as a universal binary format.
By virtualizing the bottom layer of userspace, kernel interfaces enable effortless application ISA portability, compiler backend reusability, and armor programs with Wasm's built-in control flow integrity and arbitrary code execution protection.
Furthermore, existing capability-based APIs for Wasm, such as WASI, can be implemented as a Wasm module over kernel interfaces, improving reuse, robustness, and portability through better layering.
We present an implementation of this concept for two kernels -- Linux and Zephyr -- by extending a modern Wasm engine and evaluate our system's performance on a number of sophisticated applications which can run for the first time on Wasm.
\end{abstract}





\maketitle

\section{Introduction}







WebAssembly (Wasm) \cite{WasmPldi} has emerged as a lightweight, efficient virtualization solution applicable to many domains.
As a portable low-level bytecode format with a strict formal specification~\cite{WasmSpec}, type system with machine-checked proofs~\cite{WasmMechSpec}, and high-performance implementations~\cite{Wamr} with ever-increasing levels of verification~\cite{WasmSandboxing}, Wasm provides an efficient sandboxed execution environment which can run untrusted code at near-native speeds\footnote{\difftxt{The Wasm language specification is continually advancing, incorporating various enhancements for SIMD vectorization~\cite{WasmSimd,WasmRelaxedSimd}, larger address spaces (memory64~\cite{WasmMemory64}), and finer-grained memory control (multi memory~\cite{WasmMultiMemory} and custom page sizes~\cite{WasmCustomPageSizes}).}}.
Primarily deployed in the Web today, it serves as a polyglot compilation target powering many applications such as Photoshop~\cite{niemela2021webassembly}, Unity~\cite{Unity}, and high performance libraries~\cite{WasmCrypto}.

Following its success in browsers, Wasm has also gained broad adoption in cloud and edge contexts~\cite{WasmEdgeDancer,CloudflareWasm,FastlyEdge}.
To operate in these contexts, Wasm requires a \emph{system interface}, as its core specification defines a portable bytecode ISA without any system interfaces.
Outside Web APIs, the WebAssembly System Interface (WASI)~\cite{WASI} is the only proposed standardized platform interface.
WASI is a secure, cross-platform (OS-agnostic) Wasm interface specification that couples a system interface with a new capability-based security model, enforcing filesystem isolation with pre-opened directories, network isolation with constrained sockets, and explicitly-enumerated environment variables.



Recent years have seen growing interest in extending Wasm beyond controlled cloud environments to modern cyber-physical deployments at the edge, incorporating highly-capable mobile and internet-connected embedded devices such as IoT \cite{li2022bringing,WasmIotOs,Aerogel,JitWasmEmbedded}, automotive \cite{etas-wasm, scheidl2020webassembly}, and industrial systems \cite{siemens-wasm}.
Unlike the cloud, these domains have several uniquely challenging requirements. 
Software in these domains often demands high performance and memory efficiency, operates in safety-critical physical environments, and combines components distributed by many vendors in their preferred choice of languages.
Applications are also often deployed across a wide gamut of system configurations for long deployment periods. 
Many critical software stacks are hence presently frozen in time on legacy hardware and software ecosystems that are rapidly falling behind the state of the art for safety, efficiency, and performance.

Wasm is a compelling solution for these problems, addressing efficiency, safety, and polyglot concerns.
However, the additional requirements of high portability across software platforms, long deployments, and critical legacy software are system interface concerns which Wasm explicitly does not address.
To date, no existing standard system interface for Wasm adequately addresses these challenges.
In particular, WASI's goals are somewhat \emph{misaligned} with these challenges: 
\begin{enumerate}[(1),topsep=0pt,noitemsep]
    \item As a \emph{new} portable API across many operating systems, it must be reimplemented many times;
    \item Its design exploration and evolution make it unstable and therefore unsuitable for long deployments; and
    \item Its divergence from longstanding standards like POSIX means it cannot run existing software.
\end{enumerate}
Furthermore, despite many years of standardization efforts, WASI remains an extremely simplified OS interface, lacking support for prevalent OS features like memory-mapping, processes, asynchronous I/O, and signals. 
Many applications require these features and simply cannot run on WASI (see Table.~\ref{table:wali_compiled}).
Given the many benefits of Wasm as an execution format, we believe the lack of an effective system interface is a key limiting factor to Wasm adoption in many domains.

In this paper, we propose \textit{thin kernel interfaces} for Wasm to enable a new virtualization solution for userspace applications. 
Our key insight is that operating systems' userspace syscall interfaces are a stable, de facto standard upon which thousands of desktop, server, and embedded applications have been and continue to be built.
By creating a thin Wasm interface for existing operating systems, we can easily virtualize entire software stacks from the bottom-up with little modification and run them on a diverse set of host ISAs.


\begin{figure}[t]
\centerline{\includegraphics[width=1.0\columnwidth]{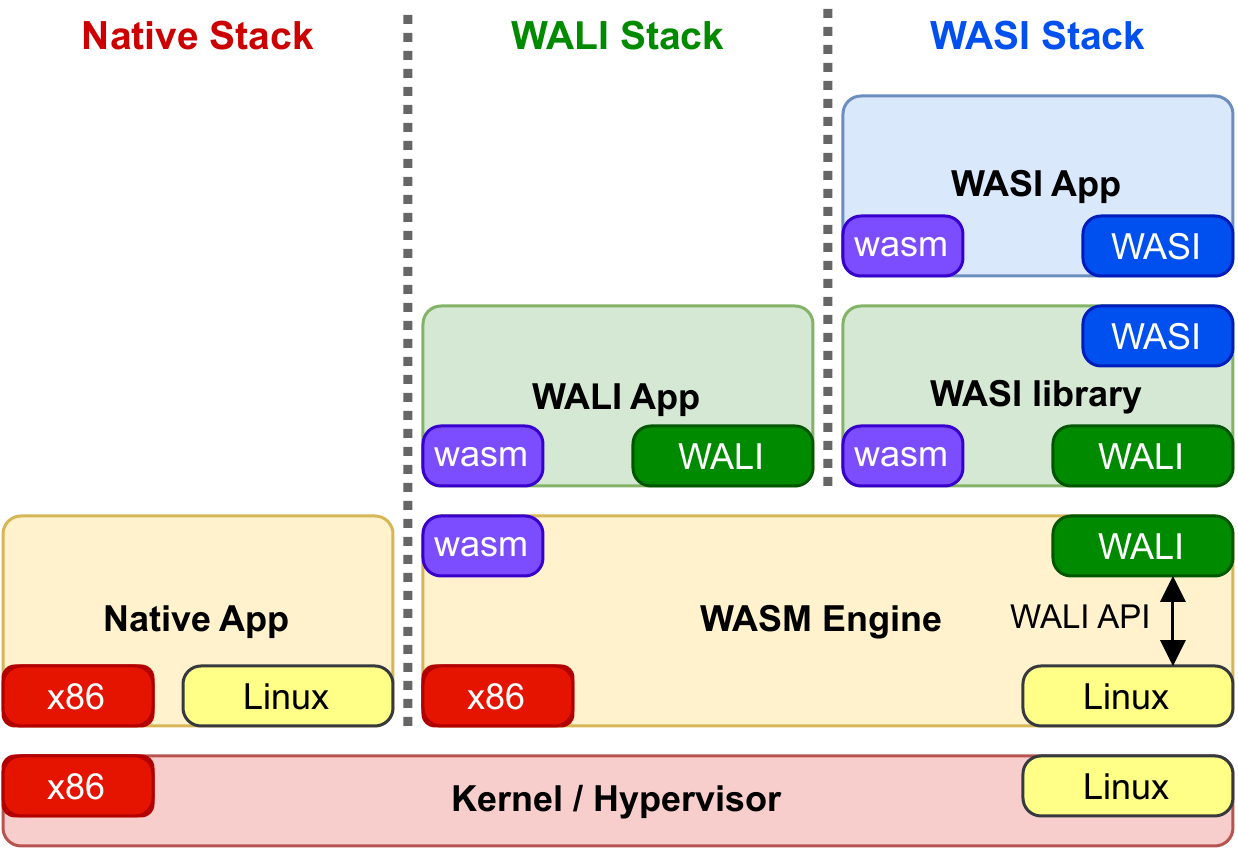}}
\caption{Linux Virtualization stack with \sysname as a foundation.}
\label{fig_wali_layering}
\end{figure}

In contrast to existing system interfaces for Wasm, thin kernel interfaces do not define an entirely new API surface against which applications need to be refactored, but rather faithfully model the existing underlying operating system. 
The bottom-up approach means that porting applications only requires recompilation with a Wasm-enabled toolchain, which can be done quite trivially as standard libraries and ABIs are already written against the OS syscall API.
Furthermore, kernel interfaces are both complementary and advantageous to higher-level APIs like WASI as a complete syscall specification allows them to be implemented as individually-sandboxed layers over the kernel interface (Fig.~\ref{fig_wali_layering}) as Wasm modules. 
Such layering makes high-level Wasm API implementations more \textbf{portable}, \textbf{safe}, and \textbf{reusable}, allowing them to work on any Wasm engine that exposes the same kernel interface.
Furthermore, the complex trusted computed base (TCB) of the Wasm engine is now greatly simplified, as high-level APIs once necessarily embedded within the engine are now \textit{decoupled} from it.

We demonstrate the feasibility of this concept with two distinct OS interfaces ---  a \textit{Linux} interface \textbf{\sysname}, and a \textit{Zephyr} RTOS interface \textbf{\zephyrsysname}. 
Beyond these two implementations, we also present a \textit{\textbf{recipe}} for Wasm kernel interface design applicable to any kernel, which addresses important architectural decisions in safely supporting common OS features like signals, processes,  threads, and memory mapping to bridge the mismatch between Wasm and a typical OS execution model.

\subsection{Motivation}
\label{sec:motivation}
\paragraph{Virtualization Requirements at the Edge}
Modern edge systems are complex pieces of software deployed on highly-customized hardware platforms.
Software virtualization in this setting has the following requirements: 

\begin{enumerate}[label=(\arabic*),itemsep=0.2em, topsep=0.2em]
    \item \textbf{High ecosystem portability}: Applications in manufacturing equipment or automotive systems are deployed across a wide array of systems and engines.
    A simple interface implementation must be able to easily port complex software stacks across many engines.
    \item \textbf{Long deployments}: Many applications are deployed for decades without modification. Supporting such deployments requires a target with a \textit{stable} set of features that is complete with low churn.
    \item \textbf{Critical legacy software stacks}: Many industries have large actively-deployed legacy codebases that cannot easily be rewritten. 
    Applications should be virtualizable "as-is", with simple update mechanisms allowing for incremental software improvements.
    \item \textbf{Efficiency}: Efficient CPU/memory usage and package size is critical in resource-constrained devices deployed in the wild. Highly reactive environments may also depend on fast application startup times.
    \item \textbf{Safety and Security}: Safety-critical physical environments such as control systems and factories demand safe execution and resource control. 
    Applications must be statically-typed, verifiable, and isolated from other co-located applications. 
    \item \textbf{Polyglot}: Applications should be easy to develop and port from a variety of programming languages and support the vast set of software libraries currently available to these languages. 
\end{enumerate}

\paragraph{Why Wasm?}
Given the stakes, safety should be paramount for a virtualization solution, especially in actively-deployed systems susceptible to remote code injection attacks.
Containerization (e.g. Docker~\cite{docker}) is too memory inefficient, and is still vulnerable to control-flow redirection and remote code execution (RCE) attacks.
Bytecode VMs often meet the aforementioned virtualization requirements, but targeting most managed runtimes (e.g. Java and CLR) restricts application development to specific families of languages, often with builtin non-determinism and garbage-collection, making them unsuitable in embedded and real-time contexts.

\difftxt{In contrast to the above solutions, Wasm offers a secure and efficient compilation target for numerous high-level languages. Armed with kernel interfaces, Wasm binaries inherently provide the following core properties by design:
\begin{enumerate}[(1),topsep=0pt,noitemsep]
    \item \textbf{Type Safety}: Wasm binaries are type-safe and statically validated prior to execution;
    \item \textbf{Memory Safety}: All Wasm memory accesses, including \texttt{NULL} and type unsafe pointer indirection, are in-memory sandboxed and runtime-enforced. Default memory initialization also prevents undefined behavior;
    \item \textbf{Non-Addressable Execution State}: Wasm's static code segments and its dynamic call/value stack are neither directly addressable nor aliasable from within the module, ruling out arbitrary code injection/execution attacks;
    \item \textbf{Control Flow Integrity (CFI)}: Wasm's structured control flow paradigm along with typed function calls and a virtualized call stack provides implicit CFI~\cite{CFI}.
    \item \textbf{ISA Portability}: The Wasm bytecode format and semantics are defined independently of hardware ISAs, allowing universal compatibility with any underlying host;
\end{enumerate}
}



\paragraph{Protecting System Software}
Real-world edge deployments often include a large exploitable attack surface of not just applications but dozens or even hundreds of susceptible background system services (e.g. remote logins, authentication, daemons) and libraries. 
By using Wasm to secure system daemons (see Table.~\ref{table:wali_compiled}), thin kernel interfaces could have immediately thwarted numerous \textit{OpenSSH}~\cite{SshCVE} vulnerability exploits, including the critical \textit{regreSSHion} bug~\cite{regreSSHionCVE} that affected millions of Linux devices worldwide.
Securing a large swath of the system software stack with Wasm in this manner, however, requires an ideal interface that supports a diverse set of OS features.

\paragraph{The Need for Complete Wasm OS Interfaces}
Wasm as a virtualization solution for edge systems is clearly promising, but high portability, long deployments, and legacy software support require the right system interface compatible with these goals.   
Standardized efforts towards this end include:
\begin{itemize}[noitemsep, topsep=0em]
    \item \textbf{WASI}~\cite{WASI}: a W3C standardization effort that aims to design a completely portable Wasm API with a strict capability security model, fine-grained access controls, a simplified filesystem, and socket virtualization.
    \item \textbf{WASIX}~\cite{Wasix}: a rogue superset of WASI, proposed by the Wasmer~\cite{Wasmer} team that adds missing POSIX functionality to jumpstart WASI development. 
\end{itemize}
As API specifications intent on integrating both OS portability and a capability-based security model, both WASI and WASIX face major design difficulties in maintaining compatibility with existing application APIs like POSIX.
For one, essential features like signals, memory-mapping, or users/groups have been eschewed due to design difficulty; as a result, applications that use these \textit{just don't work}.
Yet despite WASI's limited feature set, implementing WASI is surprisingly complex\footnote{libuvwasi~\cite{libuvwasi} for preview1 is over 6,000 lines of engine code, excluding \textit{libuv} itself or the component model which is several thousand more!} with concomitant engine requirements, and many implementations are riddled with bugs~\cite{wave-sandbox}.
Further exacerbating the problem, WASI is growing with new features for machine learning, HTTP, key-value stores, etc., as all engines engine are forced to implement these internally, in the trusted computing base.
Bugs in these implementations can compromise the inherent memory sandboxing, CFI, and RCE defense, effectively breaking Wasm.

Our approach instead adopts a layering approach to API design, targeting historically stable OS-specific syscall interfaces, and hence \textit{decoupling the security model from the feature-completeness} of Wasm as an ISA.
\difftxt{By deprioritizing OS-portability as a primary goal, kernel interface implementations remain remarkably tiny (\sysname $\approx$ 2000 LoC) compared to their higher-level API counterparts, and offer feature-completeness for running complex legacy applications, including layered WASI implementations over it.
}
\difftxt{These kernel interfaces implementations are also easily adaptable to support the handful of popular OSes by following our \textit{recipe} in Sec.~\ref{sec:beyond_linux}.}
Engines can now support a plethora of arbitrary high-level APIs for an OS in a portable, sandboxed manner above a single implementation of a stable syscall interface. 
This is especially valuable for kernels such as Zephyr that to date have \textit{no} engine with a complete WASI implementation.

\subsection{Positioning Kernel Interfaces in the Ecosystem}

Kernel interfaces hold a unique position in the Wasm ecosystem without diminishing the use-cases of existing capability-based security APIs. 
\difftxt{They enable a new ecosystem providing \textit{traditionally native} binary software stacks (e.g. managed edge systems, OS packages, consumer mobile/desktop applications, and WASI implementations) with both a viable virtualization target for safety and portability, and an opportunity to run alternative security models.}
Allowing a full set of OS system calls (e.g. Linux) raises some obvious questions:

\questionanswer{Do Kernel Interfaces Break Wasm?}{
\difftxt{No, Wasm actively facilitates support for arbitrary custom system interfaces via import sections in its language standards. 
This paper demonstrates how kernel interfaces can be exposed safely while preserving the most essential Wasm properties listed in Sec.~\ref{sec:motivation}.
Importantly, kernel interfaces do not directly tamper with the execution stack and instructions in \sysname/\zephyrsysname modules still operate within their own isolated memories (Sec.~\ref{sec:security_model}).
}
So-called ``risky'' features like \texttt{setjmp/longjmp} in C/C++ would be compiled to safe, non-local control flow with exception-handling akin to the browser, making them a toolchain concern rather than a system interface issue.
}

\questionanswer{Do Kernel Interfaces Ruin WASI/Browser Ecosystems?}{
No, kernel interfaces are \textit{not meant to replace WASI} or be directly exposed to untrusted cloud software or as browser APIs. 
WASI and browser ecosystems will continue to operate with their capability-based security models.
\difftxt{In fact, we propose no changes to their APIs or security models, which are undoubtedly beneficial for their intended cloud or Web deployment scenarios.}
We contend however that engines will be more secure if they move their WASI implementations up to Wasm and layering it over kernel interfaces (Sec.~\ref{sec:porting_effort}) \textit{without exposing the latter directly} to user applications.
}

\subsection{Contributions}

\begin{enumerate}[label=(\arabic*),itemsep=0em,topsep=0em]
\item We identify \emph{thin kernel interfaces} for Wasm as the key layering mechanism that simultaneously decouples high-level APIs (e.g. WASI) and engine evolution, improves safety by sandboxing their implementations, and allows numerous legacy applications to run for the first time on Wasm across multiple hardware ISAs.
\item We design and implement \sysname, an interface for Linux, where we \difftxt{support a critical mass of syscalls,} explore several process and signal models, and evaluate the implementation on sophisticated real-world applications.
\item We propose a \textit{recipe} for defining ISA-agnostic virtualization for OS kernels using Wasm and apply this recipe to design \zephyrsysname, an interface for Zephyr RTOS.
\item We showcase decoupling WASI from engines by compiling \texttt{libuvwasi}, a popular implementation of WASI, unmodified over \sysname with full feature-completeness.
\end{enumerate}
The \sysname and \zephyrsysname implementations are fully open sourced and available at \url{https://github.com/arjunr2/WALI}.

\section{Scoping Existing System Call Interfaces}
\label{sec:scoping_syscalls}

Operating systems feature a daunting array of hundreds of system calls, sometimes slightly different across ISAs, including some platform-specific calls.
While virtualizing every call across all platforms seems like a gargantuan task, our study of real-world applications found Linux syscalls have enough commonality in \textit{actual use} that \sysname can cover the vast majority of OS functionality in a lightweight, portable way.

\begin{figure}[t]
\centerline{\includegraphics[width=1.0\columnwidth]{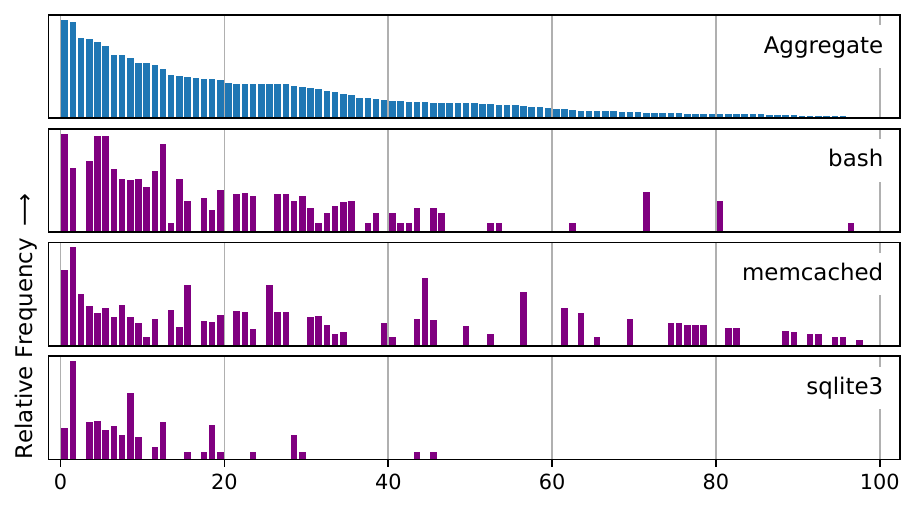}}
\caption{Log-normalized Linux syscall profile sorted by aggregate frequency; the top row shows the distribution of \textit{all invoked syscalls across all benchmarks} sorted by frequency; lower rows show the syscall frequency for each benchmark using the same ordering.}
\label{fig_app_syscall_profile}
\end{figure}

\begin{figure}[t]
\centerline{\includegraphics[width=1.0\columnwidth]{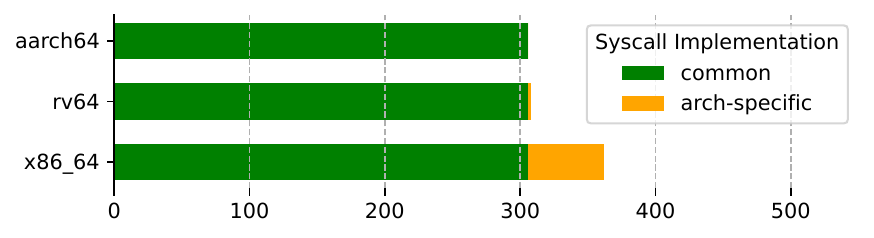}}
\caption{Similarity of Linux Syscalls across ISAs.}
\label{fig_wali_arch_syscalls}
\end{figure}

\begin{figure*}[t]
\centerline{\includegraphics[width=1.0\textwidth]{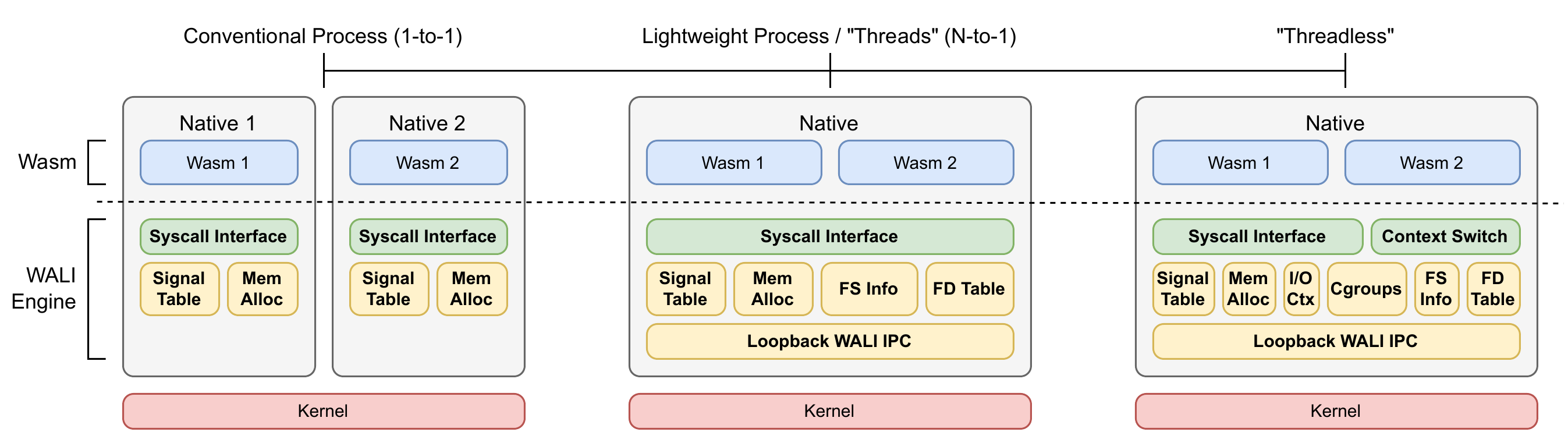}}
\caption{Process Model Spectrum for varying configurations of \textit{Native} and \textit{Wasm} processes; \sysname must implement the \textbf{bolded} components.}
\label{fig_wali_process_models}
\end{figure*}

\paragraph{Practical Prevalence of Syscalls}
We first evaluated the feasibility of syscall-based virtualization to scope our work on \sysname/\zephyrsysname by studying the number and frequency of syscalls across a variety of applications. 
Linux API usage literature reports $\approx$ 200 (out of 300-400) required syscalls as the critical point with static analysis~\cite{tsai2016study}; more recent work with dynamic analysis~\cite{Loupe} however reports a mere 148 syscalls which is consistent with our findings (Fig.~\ref{fig_app_syscall_profile}). 
This is consistent with our findings (Fig.~\ref{fig_app_syscall_profile}) where many applications use fewer than 100 unique syscalls, and the union of all applications is around 140-150 syscalls.
Similarly, Zephyr has $\approx$ 500 syscalls, but many of them target domain-specific subsystems (GNSS, SiP SVC, Auxdisplay, etc.) or toolchains. 
Thus, \sysname/\zephyrsysname only need to support a fraction of the total system call interface to run most applications faithfully.
We also identify that this subset of syscalls encompasses the core features offered by an operating system -- we postulate that most remaining syscalls can easily be auto-generated as kernel \textit{passthrough} methods in the future (see Sec.~\ref{sec:beyond_linux},~\ref{sec:discussion}).

\paragraph{Diversity Across ISAs}
Linux officially supports $\approx$ 500 syscalls but not all syscalls are available on all ISAs~ \cite{LinuxSyscalls}.
We found there is a large common core, as both Arm and RISC-V are nearly identical and largely a subset of x86-64 with a handful of differences (Fig.~\ref{fig_wali_arch_syscalls}).
These syscalls persist for backward compatibility but can often be emulated using newer, more secure alternatives (e.g., \texttt{access} with \texttt{faccessat}, all \texttt{stat} variants with \texttt{newfstatat}).
Meanwhile, Zephyr is conveniently designed to be ISA-portable across all its diverse target platforms/boards.
Thus, \sysname/\zephyrsysname can feasibly attain interface-level ISA portability will minimal effort.

\section{An Interface for Linux: \sysname}

As a ubiquitous, robust, powerful, and stable OS, we choose Linux as the primary target to demonstrate the feasibility of thin kernel interfaces.
\sysname specifies a set of approximately 150 WebAssembly host functions that can be imported into a Wasm module.
Since most Wasm runtimes support extensible host functions, \sysname implementations are relatively easy to add to engines.

\sysname is comprised primarily of syscalls, with a small number of support methods for environment/command-line parameters. 
\sysname syscalls correspond nearly 1-to-1 with native Linux syscalls and need only translate data between the virtualized syscall interface and the native Linux syscall interface and vice versa.
By design, most \sysname calls are "\textit{passthrough}", with low-overhead \textit{zero-copy} operations with appropriate translation between the Wasm and Linux memory spaces.
\sysname maintains fundamental Wasm guarantees (ISA portability, memory sandboxing, CFI, Harvard architecture) and thus disallows register accesses, stack access, and non-local gotos (\texttt{setjmp}/\texttt{longjmp}).

Considering all these components, \sysname must make fundamental important design decisions to bridge the Linux and WebAssembly execution environments --- primarily for the process/thread model, memory management model, signal handling, security model, and cross-platform support.

\subsection{Process and Thread Model}

While most operating systems support concurrency using a process/thread model, core Wasm provides no notion of processes or threads. 
This requires \sysname to present a process/thread model to Wasm applications that faithfully represents the behavior of native Linux processes in order to seamlessly interact with each other and with native processes.
Conventional Linux processes commonly interact with each other using pipes, shared memory, or signals, while threads/light-weight-processes (LWP) often share a common memory space and communicate using synchronized memory operations. 
Many syscalls additionally use \emph{process ids} (PIDs) to target processes for calls that involve signals, usage statistics, scheduling characteristics, and status updates.
To faithfully replicate these behaviors, we explore three models for \sysname along the spectrum of possible models (Fig.~\ref{fig_wali_process_models}):

\paragraph{\singlewp model}
In this (simplest) model, each \sysname process is assigned a unique native Linux process with its own PID. 
A key advantage of this model is the ease of implementation and verification: most process/thread-oriented \sysname syscalls, including \texttt{fork}, can be implemented as pass-through syscalls directly to the kernel.
This design also relieves the engine from maintaining any \sysname native process/thread state, but places the engine at the mercy of the Linux kernel for further optimizations in performance or inter-process communication.
\difftxt{
To work around Wasm's lack of a thread model, both the web and WASI support threads via replicating Wasm module instances.
This ``instance-per-thread'' model preserves distinct execution state, including separate value and call stacks\footnote{\difftxt{The design around Wasm thread support may be standardized in future with the Shared-Everything Threads proposal~\cite{WasmSharedEverythingThreads}}}, globals, and tables.
We adopt the \singlewp design with an instance-per-thread for our \sysname implementation for simplicity, and it incurs zero bookkeeping memory overhead of processes/threads within the engine.}
    
\paragraph{\multiwp model}
The \multiwp model runs multiple \sysname processes as \textit{LWPs} within a single Linux process. 
\textit{Thread-based} LWPs require virtualization of all unshared native process state, significantly increasing implementation complexity.
Luckily, the native Linux \texttt{clone} call supports a precise specification for fine-grained resource sharing with the child process, allowing \sysname implementations to optimize tradeoffs on the spectrum between conventional "processes" and "threads", e.g.:
\begin{itemize}[noitemsep, topsep=0pt]
    \item Setting \texttt{CLONE\_VM} allows the child LWPs to share the parent's virtual address space, enabling potential memory usage optimizations.
    \item Disabling \texttt{CLONE\_THREAD} makes interactions with virtual LWPs identical to conventional processes: they obtain a unique thread-group ID (TGID) and possess their own scheduling properties.
\end{itemize}
Co-locating multiple \sysname processes in one native process also allows sharing filesystem information and signals, which can allow fast inter-process communication without syscalls which may even outperform native Linux IPC.  

\paragraph{"Threadless" model}
Mode switch overheads for kernel calls are becoming more significant with recent exponential improvements in CPU and memory performance~\cite{zhong2022xrp}.
Leveraging Wasm's sandboxing, further reduction in overhead could be achieved by avoiding an LWP-backed process model in favor of a hyper-optimized process model that runs Ring-0 delegated tasks in user-space~\cite{Wasmachine}. 
We imagine that a \threadlesswp model could support context switches as fast inter-instance function calls within the Wasm engine in user-space, eliminating mode switch overheads. 
TGID-based process identification can be emulated with a \emph{dummy} native process that forwards process-based interaction to the \sysname engine, or with basic kernel support for providing raw TGID identifiers.   

\subsection{Memory Model}

Wasm module memory is a 32-bit~\footnote{With the recent standardization of the \texttt{memory64} proposal~\cite{WasmMemory64}, Wasm memories can be optionally extended to 64 bits.} byte-addressable, bounds-checked linear address-space instantiated as subset of the host process's memory space. 
Module memory declarations statically specify an initial and maximum number of (64KiB~\footnote{A recent Wasm proposal would allow custom power-of-two page sizes~\cite{WasmCustomPageSizes}.}) pages that are shareable by multiple parallel computations (i.e, \emph{threads} on WASI and \sysname), and accessible by the Wasm program using load/store instructions that generate a 33-bit index into memory.
These memory model details necessitate design choices to support \sysname syscalls that operate on data in memory or perform memory management. 
We transparently support all memory-oriented syscall operations with the following techniques: 

\paragraph{Address-Space Translation} 
For many native syscalls that accept arguments representing pointers to process memory regions, WebAssembly memory "pointers" cannot be directly forwarded to native syscalls as pointer types. 
For such \sysname syscalls, the engine must perform an \textit{address-space translation} of memory references between Wasm and native process memory.
This fast linear translation with minimal bounds-checking overhead allows most \sysname syscalls to be \emph{zero-copy}, enabling high-performance I/O directly from the sandboxed intra-process Wasm memory.

\paragraph{Layout (ABI) Conversion} 
Some native syscalls accept pointers to complex structured-typed arguments, whose byte-level layout and size of may vary across ISAs, making it impossible for WebAssembly to provide these as platform-independent zero-copy syscalls.
In such situations, \sysname must explicitly perform Wasm-to-native \texttt{struct} copies for input arguments and native-to-Wasm copies for output arguments. 
Few syscalls (<10\%) use such arguments and their sizes are usually small and fixed, imposing minimal overhead.

\paragraph{Memory Management}
\sysname allows nearly all use cases of \texttt{mmap}, \texttt{mremap}, and \texttt{munmap}, including mapping files and other resources with unconstrained address ranges.
All allocations are fully sandboxed and mapped by the \sysname implementation within the Wasm memory address space. 
Our implementation automatically grows Wasm memory for new mappings, up to the memory declaration's self-imposed limit, failing if the size grows beyond the maximum.
\footnote{
    We utilize the \texttt{MAP\_FIXED/MREMAP\_FIXED} flag to native \texttt{mmap/mremap} syscall to map pages at specific addresses in Wasm memory.
}
Subsequent unmapping with \texttt{munmap} is performed as a passthrough native syscall with normal bounds-checking.

Prior to \sysname, source-level language memory management libraries (e.g. \texttt{malloc} or garbage collection) were reliant on the \texttt{memory.grow} instruction, which required non-trivial porting effort to run on Wasm.
With \sysname's faithful memory-mapping support, sophisticated mapping strategies work unmodified using kernel interfaces, allowing these libraries to immediately target Wasm.
Such support obligates the \sysname engine to internally manage the state for allocated mapped and free memory segments.
Our implementation allows mapping a region in the engine at most once \difftxt{(which only requires a single bookkeeping variable)} for tracking the base address of the allocation pool.
Future implementations however may avoid fragmentation with more elaborate allocators -- these can implemented as Wasm modules over \sysname to reduce engine complexity and improve portability.


\subsection{Signal Model}
Both synchronous and asynchronous signal handling are critical features used by many Linux programs.
Synchronous signals are generated and delivered immediately to processes in reaction to most hardware faults, e.g. memory access faults, illegal instructions, or arithmetic exceptions.
These are easy to catch and trigger \textit{traps} in the Wasm engine for safe exception handling (e.g. \texttt{SIGFPE} for integer division-by-zero).
Asynchronous signals however are more challenging: they may be generated and delivered at any point in a process's lifespan, even while the target process is suspended, and are frequently used for software interrupts, job control, termination, or I/O.
WebAssembly, at the time of writing, has no standardized instructions for asynchronous callback operations. 
As a result, the \sysname engine must explicitly support both delivering asynchronous native signals at Wasm bytecode \emph{safepoints}~\cite{GcStackmaps} and executing user signal handlers faithfully.

\paragraph{Asynchronous Signal Handling}

\begin{figure}[t]
\centerline{\includegraphics[width=1.0\columnwidth]{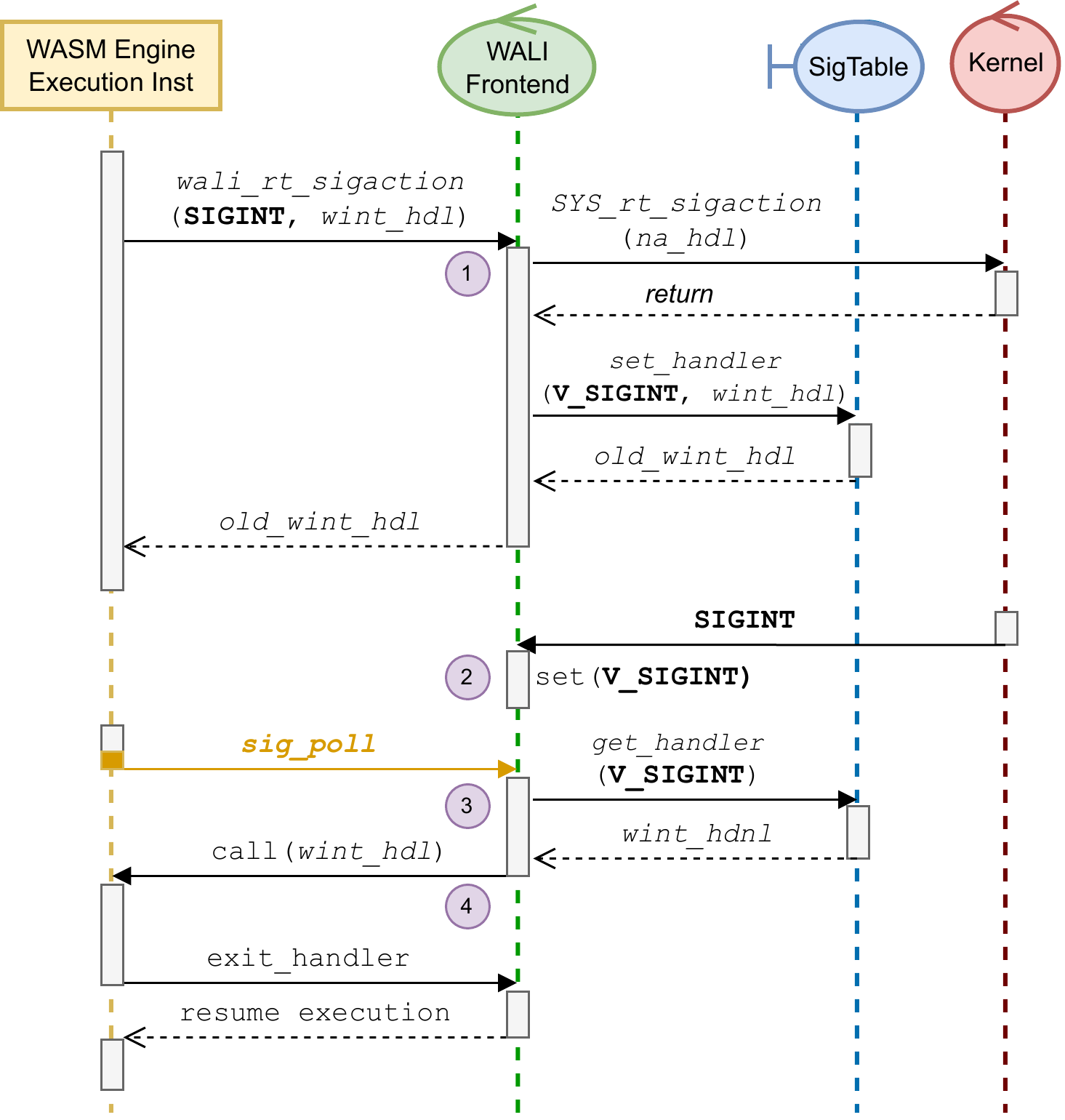}}
\caption{\sysname Asynchronous Signal Handling Sequence Diagram}
\label{fig:wali_signal_handler}
\end{figure}

\sysname engines must be capable of supporting asynchronous signal delivery, masking, and execution of application Wasm functions that handle signals, similar to that of native processes. 
To fully support asynchronous signal handling, the Wasm engine must support reentrancy where a host function calls back into the same Wasm module from which it was invoked. 
\sysname implementations leverage this capability to effectively virtualize the main stages in a Linux signal's lifecycle: signal registration, generation, delivery, and handler execution (Fig.~\ref{fig:wali_signal_handler}). 

\paragraph{(1) Signal Registration:} 
Wasm modules must be able to configure asynchronous signal callback functions, akin to native process, with  \texttt{rt\_sigaction}.
To support this, the \sysname engine internally maintains a virtual \textit{sigtable} of registered signals, mapping every Linux signal to a target callback function in the Wasm module.
When Wasm modules invoke their virtual \texttt{\sysnamelower\_rt\_sigaction} syscall, two things occur:
\begin{itemize}[noitemsep, topsep=0pt]
    \item The Wasm function pointer (index into a Wasm table) is dereferenced and registered in the \textit{sigtable}; and
    \item The native \texttt{rt\_sigaction} is called within the engine to register a \textit{native handler} for the signal that performs virtual signal generation.
\end{itemize}
The Wasm function pointer is also saved in the \textit{sigtable} to return back the old action (\texttt{old\_wint\_hdl}) to the module for future invocations of \texttt{\sysnamelower\_rt\_sigaction}.
\difftxt{The virtual \textit{sigtable} incurs a minimal bookkeeping overhead of $< 1kB$.}

\paragraph{(2) Generation:} The engine stores a bit-vector and a queue of pending signals per \sysname process to serve as the virtual signal generation mechanism.
Since we use \texttt{rt\_sigaction}, native signal generation is performed by the underlying kernel which \sysname, as a user-space interface, uses to set the signal's bit-vector element and add it to the pending queue.

\paragraph{(3) Delivery:} Generated signals remain pending and are delivered shortly after to the native process. 
However, signals may be blocked using a \textit{signal mask} (with \texttt{rt\_sigprocmask}) to prevent delivery until explicitly unblocked.
\sysname supports virtual signal blocking by maintaining a signal mask per \sysname process.
Since signal masks are recorded for each \textit{thread} and initial masks are inherited from the parent thread, for any process-model that uses the underlying \texttt{clone} syscall, \sysname can just use the Linux LWP's signal mask.
Delivered signals are then picked up during execution by any \sysname thread in the thread group as a result of native Linux's process model.

\paragraph{(4) Handler Execution:} Finally, the \sysname engine must trigger the execution of the registered virtual signal handler in Wasm post-delivery.
Since asynchronous signals do not impede execution, \sysname engines can choose to delay the signal delivery and handling to a later time.
However, arbitrary invocations of signal handlers during critical sections in the engine that modify module instance state (memory, tables, globals), execution environment state during \texttt{call/return} instructions, or internal \sysname state can break consistency guarantees of the WebAssembly execution model. 
Therefore, \sysname implementations must deliver signals at \textit{safepoints} (\texttt{sig\_poll} in Fig.~\ref{fig:wali_signal_handler}), inserted by the compiler, where the state consistency is preserved.
WebAssembly instruction boundaries are a natural location for safepoints, but frequent polling for signals impacts performance and compiler optimization opportunities.
Given reactivity is often non-critical, polling at loop headers and/or function entrypoints are shown to be effective solutions~\cite{basu2021frequent,iyer2023achieving} -- our implementation performs the former.

The \sysname engine must also be careful to avoid violations to basic signal delivery guarantees.
Pending virtual signals blocked by \texttt{rt\_sigprocmask} must not be delivered until the same is unblocked.
This is avoided with an additional safepoint immediately after the native \texttt{rt\_sigprocmask} invocation within the engine, which handles outstanding generated signals before entering the Wasm critical section. 
Additionally, when \texttt{SA\_NODEFER} flag is unset and two identical pending signals occur, a stack-based structure containing signal state can be used to defer new handler execution until the current handler execution completes. 
Reserved handler types like \texttt{SIG\_IGN}, \texttt{SIG\_DFL}, and \texttt{SIG\_ERR} require specialized support. The engine can allow these to bypass virtual signal handling entirely with direct calls to the kernel and special trap handlers to provide safe handling.


\subsection{External Parameters}

The \sysname specification includes methods for supporting external host parameters like command-line arguments and environment variables within the application sandbox.

\paragraph{Command-line Arguments} 
\sysname transparently supports transfer of command-line arguments from the host to the application.
To minimize state and increase safety in the engine, \sysname delegates the ownership of these variables to the standard library.
On startup, the standard library allocates an appropriately sized argument vector using two API methods --  \texttt{get\_argc} and \texttt{get\_argv\_len}.
Safe copying of each argument into the \sysname process is performed post-allocation using a \texttt{copy\_argv} method.
As a result, any security vulnerabilities exposed through buffer overflows during parsing remain entirely contained within the sandbox. 

\paragraph{Environment Variables}
Initialization of environment variables works similarly to command-line arguments in \sysname, where values are not inherited from the parent shell for security reasons but rather explicitly specified when invoking the engine. 
However, a subtle edge-case arises when executing programs internally invoke \texttt{execve}, which must pass virtual environment variables to the child \sysname process as opposed to the host engine. 
One solution for engines is to forward the current virtual environment as command-line arguments when invoking a \sysname binary. 
An alternative elegant engine-agnostic technique we adopt is to use a unique \textit{shared-memory segment} encoded with the \sysname process ID to store the virtual environment state before invocation of \texttt{execve}, which is picked up by the child process on startup.

\subsection{Cross-Platform Support}
Architecture-agnostic packaging of Wasm binaries is of prime importance for \sysname. 
Syscalls, however, are often non-portable and vary across architectures both in their syscall numbers and their functionality. 
\sysname addresses these challenges with the following techniques:

\paragraph{Name-bound syscalls} 
\sysname enables cross-ISA portability using \emph{name-bound syscalls} with statically defined type signatures.
This creates a clear distinction between the capabilities of the platform and the \sysname implementation, allowing the latter to trap if it cannot faithfully attempt the execution of a call. 
The set of \textit{virtual syscalls} in \sysname are thus a union of all syscalls across supported architectures, which serve as the single, complete \sysname syscall specification. 
Luckily, Linux syscalls show high commonality between platforms (Sec.~\ref{sec:scoping_syscalls}), simplifying cross-ISA portability efforts.

\paragraph{ISA-Specific Kernel Interfaces}
Syscall arguments like \texttt{kstat} and file status flags, used by all \texttt{stat}-related syscalls and file control syscalls respectively, have different byte-level representations across ISAs.
\sysname uses a dedicated representation for these arguments, and requires the host engine to perform layout conversion to-and-from ISA-specific representations to maintain execution consistency.
\texttt{ioctl} operations also may differ across ISAs; currently we only implement \sysname on \texttt{x86-64}, \texttt{aarch64}, and \texttt{riscv64}, which use identical operation values.
Fortunately, such disparities across ISAs are rare and require only a few lines of code to translate.

\subsection{Security Model}
\label{sec:security_model}

\begin{figure}[t]
\centerline{\includegraphics[width=1.0\columnwidth]{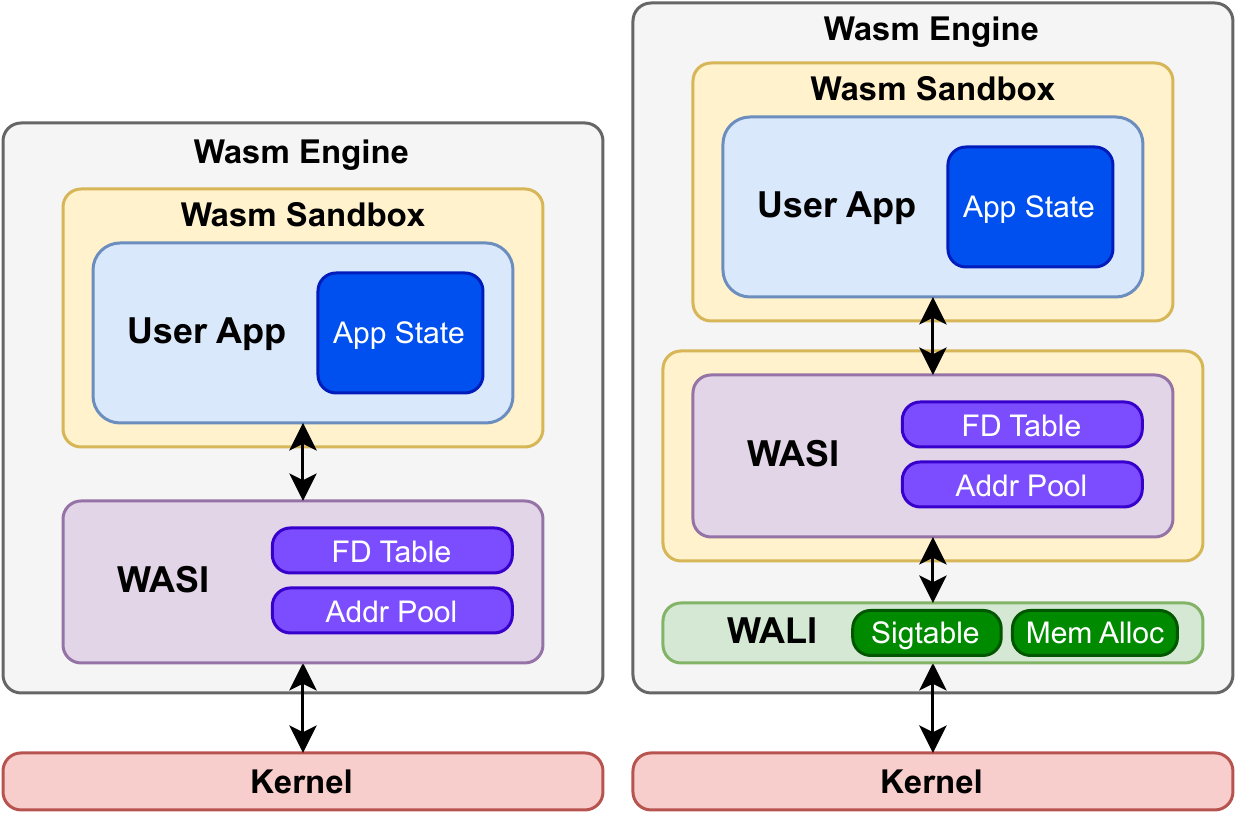}}
\caption{Minimal \sysname implementation virtualizes the WASI API.}
\label{fig_wali_state_virtualization.png}
\end{figure}

While novel security policies are at the forefront in all Wasm APIs today, this limits the engine's ability to port many existing applications and flexibility to implement new security policies.
\sysname adopts a different design philosophy to security enforcement: it maintains Wasm's inherent intra-process properties (Sec.~\ref{sec:motivation}) coupled with a purely descriptive translation to the underlying kernel (i.e. Linux).
This design pushes complex security policies to higher layers (Fig.~\ref{fig_wali_state_virtualization.png}), giving them full control in a safer and ISA portable manner.
As a result, \sysname is exceptionally thin ($\approx$2000 LoC) compared to other APIs like WASI ($\approx$6000 LoC prior to preview2), minimizing the TCB's attack surface and allowing engine developers to support numerous security policies over a single interface.
With Wasm multi-memory~\cite{WasmMultiMemory}, security models can also be provided an independent privileged memory space disjoint from the application and each other. 


\paragraph{Syscall Integrity}
\sysname binaries possess syscall integrity, explored by works such as BASTION~\cite{jelesnianski2023protect}, intrinsically with name-bound invocation, intrinsic CFI, and parameter type-checking guarantees from Wasm.
Syscalls in \sysname do not break the Wasm intra-process sandbox guarantees since:
\begin{itemize}[topsep=0pt,noitemsep]
\item \textit{All syscalls} require only \texttt{rw-} permissions to linear memory to handle data pointers, and only \texttt{r-{}-} access to table for function pointer callback arguments.
\item \textit{No syscall} manipulates the execution stack directly.
\item All instructions in \sysname binaries still only index into their private linear memories. 
\end{itemize}
\sysname binaries can also quickly be statically validated since the import section enumerates all syscalls the binary can potentially use up front, easing certification efforts.

\paragraph{Dynamic Policies}
\texttt{seccomp}~\cite{seccomp} policies are commonly used to restrict applications' syscall access capabilities for software virtualization.
\sysname does not implement \texttt{seccomp}, but rather relies on layering to implement \texttt{seccomp-like} policies completely in user-space in the engine or as Wasm modules, making existing syscall-based security work (e.g. Draco~\cite{Draco}) complimentary to \sysname.
In the long run, security policies with a constrained set of \sysname's features can serve as a simple, verifiable environment, following techniques used in security-oriented containers like Nabla~\cite{Koller2016} or gVisor~\cite{young2019true}.


\paragraph{Addressing Common Pitfalls}
Intra-process sandboxing techniques over OS abstractions are susceptible to numerous subtle security vulnerabilities~\cite{connor2020pku}.
We address these issues, with potential restrictions, in the context of \sysname :

\begin{enumerate}[label=(\arabic*), topsep=2pt, itemsep=2pt]
    \item \textbf{Filesystem Sandboxing:} Certain filesystem interfaces, notably \texttt{/proc/self/mem}, grant callers access the host process's virtual address space.
    \sysname explicitly prevents this by interposing on all \texttt{open}-like syscalls with a explicit check for the aforementioned endpoint.
    \item \textbf{Memory Mapping:} Memory-mapping calls including \texttt{mmap}, \texttt{mremap}, and \texttt{process\_vm\_\{read,write\}v} are often exploited to generate custom executable code segments.
    In \sysname, these attacks are \textit{impossible} since memory is non-executable in Wasm and all mappings are sandboxed within linear memory.
    \item \textbf{Non-Local Gotos:} C/C++ features like \texttt{setjmp/longjmp} would violate traditional Wasm CFI, performing irrevocable changes to the system stack, and are thus not supported in \sysname.
    Eventually, these features can be implemented safely transparently in toolchains as the Wasm exception-handling proposal~\cite{WasmExcept} is now part of the standard.
    \item \textbf{Signal Trampoline:} The \texttt{sigreturn} syscall is often exploited as a gadget for attacks without injecting new code~\cite{bosman2014framing}. 
    In \sysname, however, signal handler execution is fully managed within the engine, allowing us to prohibit \texttt{sigreturn} from being directly invoked from within the \sysname module user programs with a trap.
    \item \textbf{Engine Restrictions:} \sysname inherently provides maximum flexibility, but some limitations may arise from the engine's internal implementation strategies. For example, engines that use signals to trigger traps (e.g. \texttt{SIGSEGV}, \texttt{SIGFPE}) might restrict the ability of \sysname applications to override their respective handlers, or must implement chaining.
    \item \textbf{Processor-Specific Functionality:} Direct hardware access and the use of \texttt{ucontext/mcontext} are not supported in favor of ISA portability and security.
\end{enumerate}


\section{Evaluation of \sysname}
\label{sec:evaluation}

We evaluated \sysname by compiling and executing several real-world applications, build systems, and libraries. We find that this enables Wasm binaries to more easily plug into existing ecosystems with both minimal code changes and minimal API-intrinsic overhead.

\paragraph{Implementation Choices}
We evaluate \sysname with a reference implementation in the WebAssembly Micro Runtime (WAMR)~\cite{Wamr}, a popular Wasm engine written in C that supports many architectures, has extensive functionality, and has a high-performance AoT compiler in addition to an interpreter. 
For simplicity and completeness, we implement the \singlewp process model and support asynchronous signals by inserting safepoints for signal polling at loop headers for low overhead. 
\difftxt{Our \sysname implementation in WAMR currently supports \texttt{x86-64}, \texttt{aarch64}, and \texttt{riscv-64} host ISAs and was deployed successfully across a cluster of 24 diverse edge devices, including 10 resource-constrained single-board computers.}
All evaluations on \sysname below, unless otherwise specified, are collected using a fully-featured runtime on AoT compiled code.

\paragraph{Coverage}
Using our diagnostic analysis (Fig.~\ref{fig_app_syscall_profile}), we implemented the 137 most common syscalls that cover a wide range of applications compiled against \sysname to date.
The \sysname implementation is $\approx$ 2000 lines of C code, with $<$ 100 lines of platform-specific code.
We created a lightly-modified version of \texttt{musl-libc}~\cite{musl-libc} to serve as the \sysname C standard library with these notable features:
\begin{itemize}[noitemsep, topsep=0pt]
    \item full support for threads and TLS;
    \item portable versions of architecture-specific structures and flags (\texttt{kstat}, file-creation flags, \texttt{ksigaction}, etc);
    \item static linking only, since dynamic linking is not currently supported by the Wasm ecosystem.
\end{itemize}
We also build LLVM's \texttt{libc++} over \texttt{musl-libc} to target C++ applications and a new Rust target toolchain to support the Rust ecosystem \texttt{(wasm32-wali-linux-musl)}.

\begin{table}[t!]
  \centering
  \footnotesize
  \begin{tabular}{|p{5em} p{7em}|@{}P{2.5em}|@{\hspace{2pt}}P{3em}|@{\hspace{3pt}}P{2.5em}| p{4em}|}
    \hline
     & & \multicolumn{3}{c|}{\textbf{API portability}} & \textbf{Missing} \\
    \cline{3-5}
    \textbf{Codebase} & \textbf{Description} & \;\sysnamebold & \textbf{WASIX} & \textbf{WASI} & \textbf{Features} \\
    \hline 
    bash & Shell & \greencheck & \greencheck & \redx & signals \\
    lua & Interpreter & \greencheck & \greencheck & \redx & dup \\
    virgil & Compiler & \greencheck & \redx & \redx & chmod \\
    wizard & WASM Engine & \greencheck & \redx & \redx & self-host \\
    \hline
    memcached & System Daemon & \greencheck & \redx & \redx & mmap \\
    openssh & System Services & \greencheck & \redx & \redx & users \\
    \hline
    sqlite & Database & \greencheck & \redx & \redx & mremap \\
    \hline
    paho-mqtt & MQTT App & \greencheck & \greencheck & \redx & sockopt \\
    \hline
    make & CLI Tool & \greencheck & \redx & \redx & wait4 \\
    vim & CLI Tool & \greencheck & \redx & \redx & mmap \\
    wasm-inst & CLI Tool & \greencheck & \redx & \redx & sysconf \\
    \hline
    libuvwasi & WASI Lib & \greencheck & \redx & \redx & ioctl \\
    zlib & Compression Lib & \greencheck & \greencheck & \greencheck & \textemdash \\
    libevent & System Lib & \greencheck & \redx & \redx & socketpair \\ 
    libncurses & System Lib & \greencheck & \redx & \redx & pgroups \\
    openssl & Security Lib & \greencheck & \redx & \redx & ioctl \\
    \hline
    LTP & Test Harness & \greencheck & \redx & \redx & linux \\
    
    \toprule
  \end{tabular}
  \caption{Porting effort of Wasm APIs for some popular applications}
  \label{table:wali_compiled}
\end{table}
\begin{table*}[h!]
  \centering
  \small
  \begin{tabular}{|l|l l l|}
    \hline
    \textbf{Syscall} & \textbf{Overhead} & \textbf{LOC} & \textbf{State}\\
    \hline
    read & 167 ns & 4 & \textbf{N}\\
    write & 151 ns & 5 & \textbf{N}\\
    mmap & 512 ns & 30 & \textbf{Y}\\
    open & 156 ns & 4 & \textbf{N}\\
    close & 187 ns & 3 & \textbf{N}\\
    fstat & 171 ns & 4 & \textbf{N}\\
    mprotect & 120 ns & 4 & \textbf{N}\\
    pread64 & 671 ns & 4 & \textbf{N}\\
    lseek & 178 ns & 3 & \textbf{N}\\
    rt\_sigaction & 711 ns & 40 & \textbf{Y}\\
    \hline
  \end{tabular}
  \hspace{0.3em}
  \begin{tabular}{|l|l l l|}
    \hline
    \textbf{Syscall} & \textbf{Overhead} & \textbf{LOC} & \textbf{State}\\
    \hline
    stat & 112 ns & 8 & \textbf{N}\\
    futex & 141 ns & 6 & \textbf{N}\\
    rt\_sigprocmask & 114 ns & 5 & \textbf{N}\\
    getpid & 168 ns & 1 & \textbf{N}\\
    writev & 387 ns & 10 & \textbf{N}\\
    munmap & 246 ns & 12 & \textbf{Y}\\
    fcntl & 160 ns & 10 & \textbf{N}\\
    access & 202 ns & 8 & \textbf{N}\\
    recvfrom & 116 ns & 8 & \textbf{N}\\
    getuid & 151 ns & 1 & \textbf{N}\\
    \hline
  \end{tabular}
  \hspace{0.3em}
  \begin{tabular}{|l|l l l|}
    \hline
    \textbf{Syscall} & \textbf{Overhead} & \textbf{LOC} & \textbf{State}\\
    \hline
    geteuid & 123 ns & 1 & \textbf{N}\\
    poll & 128 ns & 12 & \textbf{N}\\
    getrusage & 151 ns & 5 & \textbf{N}\\
    getegid & 164 ns & 1 & \textbf{N}\\
    getgid & 165 ns & 1 & \textbf{N}\\
    lstat & 142 ns & 6 & \textbf{N}\\
    ioctl & 127 ns & 4 & \textbf{N}\\
    clone & 554873 ns & 100+ & \textbf{Y}\\
    prlimit64 & 139 ns & 5 & \textbf{N}\\
    fork & 345 ns & 1 & \textbf{N}\\
    \hline
  \end{tabular}
  \caption{\sysname implementation statistics for 30 representative syscalls guided by heuristics from Fig~\ref{fig_app_syscall_profile}, indicating the overhead (measured with VDSO-based \texttt{clock\_gettime} of \texttt{CLOCK\_MONOTONIC\_RAW}), implementation size (\textbf{LOC} --- Lines of Code), and whether the syscall is stateful. The high overhead of \texttt{clone} is notably a by-product of the engine rather than \sysname itself.
  }
  \label{table_wali_syscall_overhead}
\end{table*}

\subsection{Porting Effort}
\label{sec:porting_effort}

We collected a suite of popular Linux applications across various domains that compile using a custom \sysname clang target which is similar to the existing WASI target (Table~\ref{table:wali_compiled}). 
\difftxt{Every application compiles successfully and \textit{nearly} every one of them executes faithfully without any source code modification.
The notable exception to faithful execution arises from \textit{runtime traps} caused by failed Wasm \texttt{call\_indirect} signature checks, which occur in C applications that perform function invocation with incompatible function pointer types (e.g. bash), which is actually undefined behavior.
Interestingly, this outcome reflects a positive aspect of \sysname’s porting process, as it can help expose type safety bugs that may be latent in low-level system software.}

Surprisingly, our \sysname-enabled LLVM toolchain also seamlessly integrates into complex build systems. 
For example, Linux allows registering interpreters for custom binary formats, enabling \sysname \texttt{.wasm} files to be directly executable.
This allows many build scripts to be used directly without modification\footnote{Amusingly, the \textit{bash} build executes intermediate \sysname binaries to configure certain parameters (e.g. pipe size capacity), which run transparently without modification. This is quite common in many complex builds.}.   
The custom interpreter mechanism allowed building the entire \texttt{libuvwasi} implementation unmodified (passing all tests) and many of the currently supported syscall tests in Linux Test Project (LTP)~\cite{linux-test-project} along with their test harnesses, which uses complex signalling and shared memory for job control. 

\subsection{Intrinsic Costs}
The performance of a \sysname implementation is highly dependent on the underlying Wasm engine, which can vary drastically in performance based on how quickly it executes Wasm bytecode via interpretation or compilation~\cite{FastWasmInterp}.
While bytecode execution speed is important, it is independent of the \textit{intrinsic cost} of using \sysname that does not scale proportionally with improvements in Wasm runtimes.
Our experiments with our prototype implementation on WAMR shed light on this overhead, which is mostly independent of Wasm engine.
Experiments were run on a 11th Gen Intel Core i7-1185G7 machine (\texttt{x86-64}).
However, since architecture-specific code is minimal and infrequently executed, the intrinsic cost measured here is fairly consistent across ISAs on macrobenchmarks.

\begin{figure}[t]
\centerline{\includegraphics[width=1.0\columnwidth]{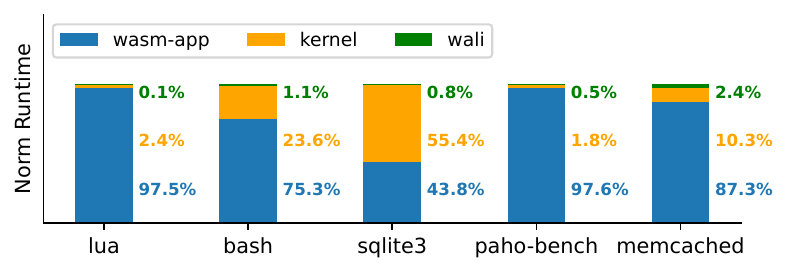}}
\vspace{-1em}
\caption{Runtime breakdown of \sysname across system stack.}
\label{fig_wali_macrobench_overhead}
\end{figure}

\paragraph{Syscall Interface}
Most \sysname syscalls require under 10 lines of code to implement -- mostly performing basic address-space translation
\footnote{Calls like \texttt{rt\_sigaction} and \texttt{mmap} typically need extra instructions to manage internal state for signal handling and memory allocation respectively, incurring a higher cost. These calls are the exception, not the norm.} 
-- and have an absolute overhead vs native syscalls in the order of a \textit{few hundred nanoseconds} (Table~\ref{table_wali_syscall_overhead}).
To put these overheads into context, less than 1\% of the execution time is typically spent in the \sysname interface, which is negligible compared to the inherent overheads of the Wasm app or kernel time (Fig.~\ref{fig_wali_macrobench_overhead}).
\difftxt{The \texttt{memcached} benchmark incurs a slightly higher overhead of 2.4\% from \sysname -- we attribute this to extensive multithreading employed by the benchmark.}

The \texttt{clone} syscall for spawning threads is a clear outlier, which adds about 500$\mu s$ of overhead. This is however not an API-intrinsic cost of \sysname, but rather that of the internal implementation of WAMR's thread manager which creates an entirely new copy of the Wasm module's execution environment for each new thread.
This cost for \sysname can be made significantly cheaper with various runtime optimizations -- e.g, the Wasmtime~\cite{Wasmtime} engine has optimized instance creation heavily through lazy loading and copy-on-write paging optimizations, resulting in overheads as low as at \textit{5$\mu s$}.
Despite this, our experience shows that most applications for our intended use-cases are not critically affected by this overhead since \texttt{clone} often occur mostly during initialization and is relatively infrequent.

\begin{table}
  \centering
  \small
  \begin{tabular}{l l l l}
    \hline
    \textbf{App} & \textbf{Loop (\%)} & \textbf{Function (\%)} & \textbf{All (\%)} \\
    \hline
    bash & 7.1 & 10.0 & 187.0 \\
    lua & 4.1 & 2.8 & 100.3 \\
    sqlite3 & 11.3 & 5.2 & 164.2 \\
    paho-bench & 0.5 & 1.1 & 17.8\\
    \hline
  \end{tabular}
  \caption{Cost of polling for asynchronous signal handling in \sysname with different safe-point insertion schemes --  \textbf{Loop}: after loop bytecode, \textbf{Func}: start of every function, \textbf{All}: after every instruction}
  \label{table_wali_signal_overhead}
\end{table}

\begin{figure*}[t]
    \centering
    \begin{subfigure}[t]{0.24\textwidth}
        \centering
        \includegraphics[width=\textwidth]{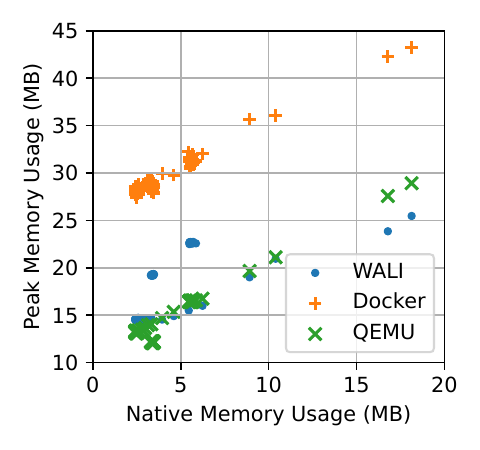}
        \vspace{-2em}
        \caption{Memory -- Combined}
        \label{fig:overhead-memory}
    \end{subfigure}
    \begin{subfigure}[t]{0.24\textwidth}
        \centering
        \includegraphics[width=\textwidth]{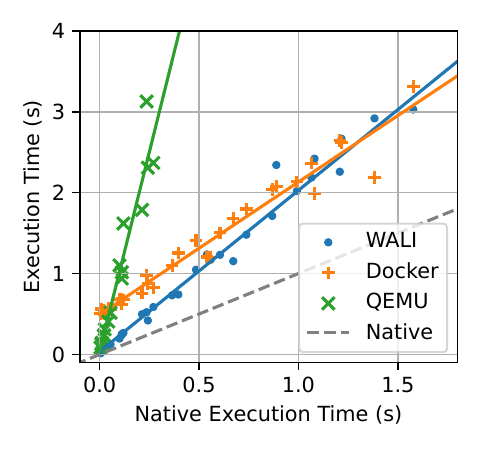}
        \vspace{-2em}
        \caption{Runtime -- Lua}
        \label{fig:overhead-runtime-lua}
    \end{subfigure}
    \begin{subfigure}[t]{0.24\textwidth}
        \centering
        \includegraphics[width=\textwidth]{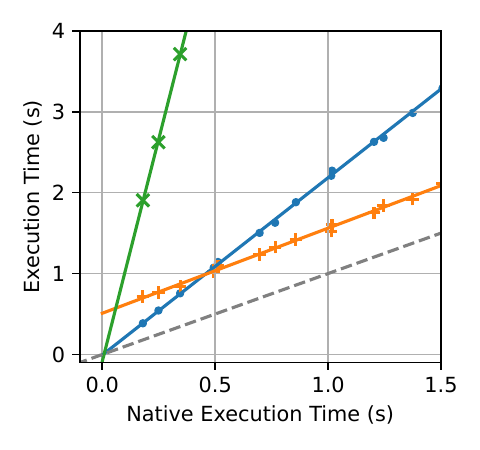}
        \vspace{-2em}
        \caption{Runtime -- Bash}
        \label{fig:overhead-runtime-bash}
    \end{subfigure}
    \begin{subfigure}[t]{0.24\textwidth}
        \centering
        \includegraphics[width=\textwidth]{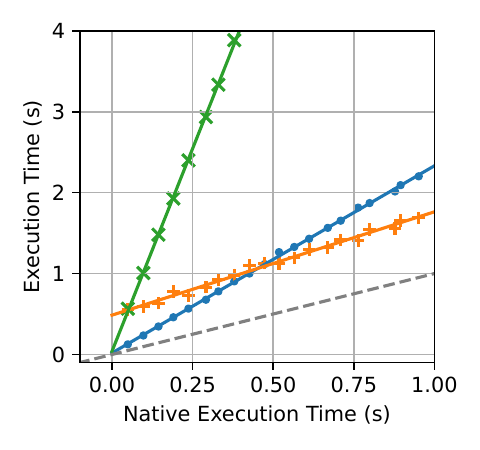}
        \vspace{-2em}
        \caption{Runtime -- Sqlite}
        \label{fig:overhead-runtime-sqlite}
    \end{subfigure}
    \vspace{-0.5em}
    \caption{Peak memory (\ref{fig:overhead-memory}) and execution time (including startup time) (\ref{fig:overhead-runtime-lua}-\ref{fig:overhead-runtime-sqlite}) comparison for Lua, Bash, and Sqlite benchmarks showing each virtualization method's efficiency versus its native counterpart; all three benchmarks are combined into a single plot for peak memory.}
    \label{fig:overhead}
\end{figure*}

\paragraph{Asynchronous Signal Polling}

The number of executed safepoints plays a critical role in execution overhead. 
Unsurprisingly, we find that polling after every instruction is prohibitively expensive--at least 10x slower than polling at loops or functions (Table~\ref{table_wali_signal_overhead}).
The latter two are comparable, typically incurring under 10\% slowdown over \sysname without signal polling. 
Both are also reasonable choices in practice -- the function scheme may favor compiler optimizations better while the loop scheme may enable more reactive signal handling for large functions.

\subsection{Extrinsic Costs: Virtualization Overhead}
\label{sec:extrinsic-costs}

\difftxt{
To put \sysname into perspective with some existing virtualization technologies, we evaluate the performance and memory usage of our \sysname implementation against two other systems --- Docker~\cite{merkel2014docker} for \textit{OS interface virtualization} and QEMU~\cite{QEMU}, specifically without KVM~\cite{kivity2007kvm}, for \textit{ISA virtualization}.
We exclude KVM-based solutions (including Firecracker~\cite{firecracker}) since KVM utilizes hardware-assisted virtualization, diverging from pure ISA virtualization that we aim to evaluate. 
We select three popular Linux applications in edge systems -- \emph{bash}, \emph{lua}, and \emph{sqlite} -- to compare and contrast the optimal use cases for \sysname with the above virtualization mechanisms.
}



\paragraph{WebAssembly Runtime Overhead}

For CPU-bound applications, the overhead of \sysname versus native applications is dominated by the Wasm engine.
A plethora of works have studied both execution and memory overheads of Wasm ~\cite{EmpoweringWasm,StandaloneWasmRt,NonWebWasmEval,titzer2024whose}, including complex optimization techniques for startup \cite{WizerEval} and bounds-checking~\cite{LeapsAndBounds,yedidia2024lightweight}, which is orthogonal to this work.
\difftxt{To provide a baseline, an analysis in 2023 of Wasm runtime performance~\cite{wasm2023-performance} shows a median slowdown (on performance-focused Wasm engines) of 2.32 times over native execution without SIMD~\cite{WasmSimd} and Tail Call~\cite{WasmTailCall} language extensions.}

\paragraph{Memory}

While peak memory utilization scales similarly for all virtualization solutions, base memory utilization can vary drastically (Fig.~\ref{fig:overhead-memory}).
Unlike \sysname, which only virtualizes the target application, Docker containers incur a high base overhead ($\approx$30 MB) to support intermediate layers for storage drivers and isolated software libraries.
On the other hand, QEMU maintains a low overhead using a several optimizations -- lazy allocation, balloon drivers, and KVM virtualization -- leading to comparable results to \sysname for small applications.

\paragraph{Execution Time}

Fig. \ref{fig:overhead-runtime-lua}-\ref{fig:overhead-runtime-sqlite} compares the composite execution time of \sysname, Docker, and QEMU.
As expected of emulators, QEMU is an order of magnitude slower than Docker, which executes at near-native speed directly on the CPU.
While our \sysname implementation's runtime performance (slope of the line) is nearly $2x$ slower than native code and Docker on average, the startup time is only a few milliseconds as opposed to nearly half a second for containers, which requires instantiation of internal layers and \texttt{namespace} isolation.
We observe a \textit{cross-over point} for each application based on start-up time and relative overheads, before which \sysname is faster than Docker.
Applications with short-lived execution or those like \emph{lua}, which execute up to 60\% slower than native in Docker due to frequent memory allocation requests, are hence good candidates for using \sysname as a virtualization solution.

\paragraph{Summary}
\difftxt{\sysname strikes a middle ground in memory and execution overhead between emulation and container technologies, providing the best of both worlds: rapid startup times comparable to emulation and faster runtime performance comparable to containers.
Furthermore, non-addressable execution state, ISA-portability, and CFI are guaranteed by Wasm's execution model, providing security benefits beyond default containerization techniques.
While Wasm, as a relatively new execution platform, still incurs high runtime overheads (consistent with state-of-the-art analysis), we envision that increased proposal standardization~\cite{WasmSimd,WasmRelaxedSimd,WasmTailCall,WasmGc} and compiler/engine enhancements~\cite{yedidia2024lightweight,titzer2024whose} will soon close its performance gap with native execution.
This will further expand \sysname's reach as a feasible virtualization technique for compute-intensive workloads. 
}
\section{Kernel Interfaces Beyond Linux}
\label{sec:beyond_linux}

Our initial design work to fit \sysname into the Wasm execution model faced core challenges such as memory-mapping, process/threads, filesystems, networking etc. that are applicable to most other kernels on modern hardware.
In particular, we formulate a \textit{recipe} for designing thin kernel interfaces for beyond Linux, which entails the following:
\begin{enumerate}[label=(\arabic*),itemsep=0em,topsep=0em]
\item\label{en:autogen-start} Enumerate and name-bind all OS calls that perform hardware-aided privilege escalation (e.g. \texttt{syscalls}).
\item Sandbox (address-space translation + bounds-check) all memory addresses passed between application and OS.
\item\label{en:autogen-end} Encode ISA-portable struct layouts and translate to/from hardware ISA struct layouts at syscall boundaries.
\item  Map the native process model onto Wasm with sandboxed processes and/or multiple threads.
\item Map kernel memory management primitives safely onto Wasm's linear memory model.
\item Map asynchronous OS interactions (e.g. signal delivery/handling, RPCs) onto synchronous Wasm interactions at safepoints.
\end{enumerate}
This concrete recipe allowed us to easily \textit{auto-generate} a majority (>85\%) of the \sysname implementation since most calls are passthrough, only requiring steps \ref{en:autogen-start}-\ref{en:autogen-end}.
Our brief investigation shows this design methodology is applicable to syscall interfaces in most operating systems.

\subsection{Applying our Recipe to Zephyr RTOS: \zephyrsysname}

To validate our recipe, we perform a small prototype implementation of our recipe for a specialized RTOS, 
Zephyr, used extensively for IoT~\cite{lee2018implementation,raymundo2018performance}, automotive~\cite{de2023study}, and industrial automation~\cite{hee2021embedded} domains, in WAMR.
Zephyr's syscall interface ($\approx$ 520 syscalls) is already ISA-portable (Sec. \ref{sec:scoping_syscalls}), and the compiler parses and creates an encoding of all syscalls at compile-time.
We extract this encoding and use it for auto-generating the implementation in WAMR, much like in \sysname. Our \zephyrsysname implementation is $\approx$ 4200 LOC.

We develop a toolchain around \texttt{picolibc}, an OS independent libc for embedded systems, and utilize \zephyrsysname to create hooks around the required I/O and filesystem calls.
In addition to simple tests, we successfully compile a \textit{Lua} binary toolchain and deploy it on a Nucleo-F767ZI ARM microcontroller board (384 kB SRAM) running Zephyr.
Given WAMR is yet to fully support WASI in Zephyr, we see this as a major step towards allowing engines to target an OS like Zephyr.

\subsection{Exploration of Other Mainstream Kernels}

\paragraph{Unix-Like Kernels} 
Most Unix-like kernels such as FreeBSD, HP-UX, and Illumos show striking similarities to Linux in their syscall set, and can follow the same design strategy as \sysname.
MacOS X, based on the Darwin kernel, marries a UNIX-style syscall set based on BSD with the Mach kernel syscalls, and implements POSIX compatibility via configurability in libc.
While enabling a modular approach with IPC for finer-grained safety and sharing, Mach still uses \sysname's core foundations albeit needing careful delineation for a kernel interface --  an orthogonal program to that of \sysname.

\paragraph{Windows}
The Windows kernel's syscalls evolve more rapidly than those of other systems, often being completely renumbered from one release to the next, with portability typically handled at the DLL level. 
While a dedicated Windows interface would likely require a stable, restricted, and portable feature subset such as the Drawbridge~\cite{porter2011rethinking} ABI, Windows systems can presently leverage WSL2~\cite{singh2020exploring} to run \sysname applications unmodified.

\section{Discussion and Future Outlook}
\label{sec:discussion}


Thin-kernel interfaces unlock a number of fruitful future directions by augmenting unmodified existing software with portability and sandboxing.
We consider a number of future directions.

\paragraph{Accelerating WASI development and adoption}

The proliferation of WASI (especially \textit{preview2}) is critically bottle-necked by engine implementation effort, with fully-featured support on only one engine (Wasmtime) and via a polyfill to JavaScript (\texttt{jco}).
Prior to kernel interfaces, WASI's complex security model was \emph{necessarily} part of the engine implementation.
Now, with WASI decoupled from from engine development, a new standardized reference implementation could be deployed as a Wasm module that uses \sysname and run on any engine.
This portability greatly accelerates the evolution and adoption of WASI on new platforms.

\paragraph{Robustness by ecosystem modularization}
Typical WASI implementations themselves contain many thousands of lines of code.
Vulnerabilities in \emph{any} of this code could compromise the memory safety of Wasm and indeed, of the entire process.
In contrast, \sysname's \emph{thin} syscall interface layer pushes more responsibility outside the trusted runtime system, reducing engine implementation complexity, increasing API stability, and sandboxing higher level APIs above the engine.

\paragraph{Portable Packaging of Linux Distributions}
Linux distributions offer pre-compiled packages for specific ISAs, which are compact, stable, and negate the need for complex build environments.
To achieve ISA-level portability, fat binaries and multi-arch Docker images have fallen short due to unwieldy disk space overhead. 
In contrast, \sysname executables are both compact and portable across CPU architectures; this raises the exciting prospect that Linux distributions could ultimately achieve full ISA portability with Wasm.

\paragraph{Full-Stack Software Verification}
Verification of native binaries~\cite{SyscallVerification} underpinned by formally-specified instruction sets~\cite{ArmSemantics,ArmSemanticsPopl} and syscall semantics~\cite{TransValidOS} have recently show great promise.
Similar efforts directed towards Wasm with machine-checked proofs of modules~\cite{IrisWasm}, fully-verified kernels~\cite{SEL4}, compilers~\cite{CompCert}, and libraries~\cite{Hacl} show promise to achieve the holy grail of verification: a tower of proofs to certify a program's entire software stack.

\paragraph{Improving Language Targetability}
We maintain that Wasm is an abstraction over hardware, rather than a specific security model for systems.
Programming languages that target Wasm should enjoy memory safety but allow full feature-completeness, which often include low-level system calls.
As more languages target Wasm, they cannot remain at the mercy of only what Web APIs or WASI allow; there must be flexibility to allow custom security layers higher in the stack to define abstractions that make the low-level interface usable, convenient, and safer.


\paragraph{Expansion and Interposition of Syscalls}
\sysname already implements most ``high-importance'' Linux syscalls~\cite{Loupe} including core OS features that require a custom bridge to Wasm. 
Our analysis of the remaining Linux API suggests that most can be added as simple pass-through calls (Sec.~\ref{sec:beyond_linux}).
Additionally, calls through Wasm can easily be interposed on by libraries that log, restrict, profile, or fault-inject.
Unlike native syscalls which are specified by a runtime syscall number, Wasm syscalls are bound by name, allowing uniform ISA-agnostic static and dynamic policies in the future.
Many tools aimed at enhancing security at the syscall layer, e.g. Nabla~\cite{Koller2016}, gVisor~\cite{young2019true}, \texttt{seccomp}~\cite{seccomp} and Draco~\cite{Draco}, are hence complementary to this work to enable a restricted subset of secure interfaces.

\section{Related Work}

We organize our discussion of virtualization technologies into four broad areas: (1) Emulators, (2) Hypervisors, (3) OS interface virtualization, and (4) Language virtualization. 

\paragraph{Emulators}
ISA emulators provide a mechanism for virtualizing an entire system stack including hardware, operating system, and application. Popular solutions like QEMU~\cite{QEMU} and Bochs~\cite{lawton1996bochs} have sparked further research into emulator performance optimizations~\cite{QemuOptimization} for niche use-cases~\cite{ QemuInfiniBandOptimization}, which is currently an obstacle to widespread adoption.
These are mostly used as prototyping tools, unless KVM~\cite{kivity2007kvm} is used when ISA emulation is unnecessary, since most binaries can run anywhere as-is, but unlike \sysname, is challenging to extend to high-performance resource-constrained systems. 

\paragraph{Hypervisors}
Hypervisor technology virtualizes the guest operating system kernel with
bare-metal (type-1) hypervisors (e.g. vSphere~\cite{guthrie2013vmware}, Xen~\cite{chisnall2008definitive}) used in cloud settings and hosted (type-2) hypervisors (e.g. Fusion~\cite{dowty2009gpu}) are used by end users. 
Hyper-V~\cite{HyperV} and KVM are common in-built type-1 hypervisors commonly leveraged in modern OSes for high performance virtualization (e.g. WSL2~\cite{singh2020exploring}, Firecracker~\cite{firecracker}).
In embedded domains, real-time hypervisors~\cite{BluevisorHypervisor,pan2018predictable,patel2015embedded} are promising for cost reduction and improved resource utilization, with bolstered security from using ARM TrustZone~\cite{pinto2016towards}.
Similar approaches that leverage hardware techniques for lightweight sandboxing~\cite{ford2008vx32} can also enable WebAssembly performance improvements.

\paragraph{OS Interface Virtualization}
Wine~\cite{WineWindows} offers a system compatibility layer for Windows applications to run on Linux but lacks any security advantages.
Containers technologies like LXC~\cite{LXC} and Docker~\cite{merkel2014docker} are popular in cloud ecosystems for high-performance isolation that uses \texttt{namespace} and \texttt{cgroups} to control resources and isolate applications.
\cite{DockerPerf} studied Docker performance in detail, reporting between 10\% and 30\% overhead for disk I/O and 5-10\% overhead when enforcing CPU quotas. 
Optimizing resource isolation~\cite{WangContainerIsolation}, startup time ~\cite{Catalyzer,manco2017my}, and heterogeneous platforms~\cite{HeterogeneousContainers} has also been a large focus of the container ecosystem for niche use-cases.
Security oriented container such as Nabla containers~\cite{Koller2016} and gVisor~\cite{young2019true} can also be mimiced via security models over \sysname/\zephyrsysname.
Wasm-based virtualization, however, provides CFI and RCE protection, along with ISA portability; advantages not currently available with containers.

\paragraph{Application Virtualization}
In the same vein as Wasm, numerous languages like Java, Javascript, Python, and .NET offer application-level virtualization. 
Browsix~\cite{Browsix} was the first POSIX-like API for in-browser Javascript applications, emulating filesystem and sockets but pays a high performance inefficiency penalty. 
\cite{stilkerich2006osek} proposed a Java OSEK interface for embedded devices and \cite{yamauchi2006writing} a Java-based device driver virtualization, but both possess non-determinism and high memory overheads for the edge.
.NET is effective in the cloud~\cite{vecchiola2009aneka} but unsuitable in edge contexts~\cite{lutz2003c} for similar reasons.
In the Wasm ecosystem, research efforts beyond WASI(X)~\cite{WasmEdgeDancer, WasmIotOs, WasmIotOs, SledgeWasm} are directed towards designing effective edge platforms and techniques to improve security for the Wasm ecosystem~\cite{WasmAvengers,SELWasm,EverythingOldIsNewAgain}, which are complementary to kernel interfaces.
Native Client (and PNaCL)~\cite{yee2010native,donovan2010pnacl} preceded Wasm for browser sandboxing, but LLVM IR is unstable, lacks full ISA portability, and contains non-deterministic behavior.
With growing interest in deeply embedded~\cite{wallentowitz2022potential,TwineWasm} Wasm runtimes, imminent domain-specific APIs will benefit from being virtualized over \sysname.

\nocite{GuestSafepoints}
\nocite{Wasix}
\nocite{DockerReview}
\nocite{QuantitativeContainer}
\nocite{ContainerStateOfTheArt}
\nocite{SolarisZones}
\nocite{WasmAvengers}
\nocite{SELWasm}

\section{Conclusion}

This paper introduced the first thin kernel interfaces for Wasm which allow high-level API evolution to be decoupled from engine and bytecode evolution, improving security, robustness, and feature-completeness through layering.
We show a repeatable recipe across diverse kernels, with examples \sysname/\zephyrsysname, that offers a strategy for developing complete, simple, and efficient thin kernel interfaces.
Wasm engines can now more easily focus on their strengths: running bytecode fast and safely exposing thin kernel interfaces, while higher-level software layers abstract over them with new security models. 
We believe this will unlock Wasm's portable software-defined ISA to expand beyond the Web or Cloud, supporting new low-level ecosystems while improving the development, distribution, and adoption of both WASI and new high-level APIs as Wasm code.


\bibliographystyle{ACM-Reference-Format}
\bibliography{references}

\clearpage

%

\appendix
\section{Artifact Appendix} 

\begin{table*}[ht!]
  \small
  \begin{tabular}{|l|l|l|}
    \hline
    \textbf{Artifact} & \textbf{URL} & \textbf{DOI/Hash} \\
    \hline
    Zenodo Virtual Machine Distribution & \texttt{\url{https://doi.org/10.5281/zenodo.14790613}} & \texttt{10.5281/zenodo.14790613}\\
    \hline
    \sysname Code Repo & \texttt{\url{https://github.com/arjunr2/WALI}} & \texttt{1e22d2e} \\
    \hline
    \sysname Experiments Repo & \texttt{\url{https://github.com/arjunr2/wali-eurosys25-data}} & \texttt{ac83120} \\
    \hline
  \end{tabular}
  \caption{List of relevant URLs for artifacts along with their commit hashes}
  \label{table:artifact-urls}
\end{table*}

\subsection{Abstract}
\textit{The artifact accompanying this paper provides public repositories for an Ubuntu 22.04 Virtual Machine and relevant code for a full \sysname ecosystem.
This serves both as a means of reproducing claims made in the paper as well as a playground to further explore the capabilities of \sysname.
This artifact excludes \zephyrsysname in the image, due to strict hardware requirements for reproducibility -- however, code artifact for \zephyrsysname is made available on GitHub.
}
\subsection{Description \& Requirements}

\subsubsection{How to access}
The experimental setup for the artifacts is contained within a Ubuntu 22.04 VM, which is made available as a public repository on \textit{Zenodo} (see Table.~\ref{table:artifact-urls} for URLs).
The repositories for \sysname code and the \sysname experiments listed in Table.~\ref{table:artifact-urls} are also publicly available on Github and \textbf{already cloned/setup} on the VM for ease of reuse.
The VM disk image is shipped in QCOW2 format its XML definition. The login information is below:
\begin{itemize}[noitemsep]
    \item Username: \texttt{evaluator} (has \texttt{sudo} access)
    \item Password: \texttt{webassembly}
\end{itemize}


\subsubsection{Hardware dependencies}
The VM should run on effectively any hardware that can support virtualization.
We recommend a system with hardware-assisted virtualization for performance reasons, with at least 8-16 cores of CPU and 8-16 GB of RAM for the VM.
The image itself consumes 26 GB of physical disk.

\subsubsection{Software dependencies} 
We recommend a Linux distribution with QEMU/KVM support to run the VM, managed with \texttt{virsh} tools.
VirtualBox may alternatively also be used, but requires converting the QCOW2 image to VDI (see \url{https://gist.github.com/mamonu/671038b09f5ae9e034e8}). 
The numerous software dependencies for the project have all been already added in the VM.

\subsubsection{Benchmarks} 
The benchmarks in Table.~\ref{table:wali_compiled} are all available in the \sysname codebase, mostly as submodules, under \texttt{applications}. 
Note, the binary \texttt{mqtt-app} is sometimes aliased as \texttt{paho-bench} in the paper.

\subsection{Set-up}

\begin{table}[ht!]
  \small
  \begin{tabularx}{\columnwidth}{|X|X|}
    \hline
    \textbf{Tar File} & \textbf{Description} \\
    \hline
    \texttt{applications-artifact} & A wide suite of applications that have been ported over \sysname, including those in Table.~\ref{table:wali_compiled} \\
    \hline
    \texttt{llvm-build} & A compressed build of LLVM to reduce disk size and save on time to build LLVM \\
    \hline
    \texttt{virt} & Miscellaneous binaries for \sysname/QEMU (\texttt{iwasm}, \texttt{qemu}, \texttt{wamrc}) for experiments\\
    \hline
    \texttt{sysroot} & (Optional) A build of \texttt{wali-musl} libc \\
    \hline
  \end{tabularx}
  \caption{List of packaged software under \texttt{artifacts} directory in the VM}
  \vspace{-10pt}
  \label{table:artifact-zips}
\end{table}

The software environment should be mostly ready to go by default on the VM + XML definition on Zenodo. The XML file defines 8 CPUs and 8 GB of RAM -- modify it as needed based on physical system constraints. You can extract and create the VM using these commands in Linux:
\begin{enumerate}[noitemsep]
    \item \textit{tar --sparse -xvf vm-ubuntu22.04-wali-eurosys25.tar.gz}
    \item \textit{sudo mv u22.04-eurosys.qcow2 /var/lib/libvirt/images}
    \item \textit{virsh define u22.04-eurosys.xml}
\end{enumerate}

\subsubsection{VM Operation}
The following are helpful commands to operate the VM with \texttt{virsh}:

\begin{itemize}
    \item \textbf{Start VM}: \texttt{virsh start u22.04-eurosys}
    \item \textbf{Shutoff VM}: \texttt{virsh destroy u22.04-eurosys}
    \item \textbf{Connecting to VM}: This can be done in 3 ways:
    \begin{itemize}[noitemsep, topsep=0pt]
        \item Console: \texttt{virsh console u22.04-eurosys}
        \item GUI: \texttt{virt-viewer u22.04-eurosys}
        \item SSH to IP address of VM, obtained from console
    \end{itemize}
\end{itemize}

\subsubsection{VM Directories} The VM should contain three directories (including repos from Table.~\ref{table:artifact-urls}):
\begin{itemize}
    \item \texttt{WALI}: The cloned repo of \sysname source code
    \item \texttt{wali-eurosys25-data}: The cloned experiments repo of \sysname, with packaged data used in evaluation
    \item \texttt{artifacts}: Compressed directory containing a variety of builds, described in Table.~\ref{table:artifact-zips} for quick setup and reducing large build size/times.
\end{itemize}

Many files have been compressed to minimize VM image size, so some setup steps below are required to perform.

\paragraph{WALI}
The \texttt{README.md} in the \sysname code repo has a setup and information guide if you intend to rebuild everything or need details. 
We have however provided everything built --- the \sysname runtime \texttt{iwasm}, the \texttt{sysroot} in \texttt{wali-musl}, and \texttt{llvm-project/build}.

The \texttt{llvm-build.tar.gz} is prepackaged under \textit{artifacts} and can be extracted with \textit{tar -xvf artifacts/llvm-build.tar.gz -C WALI/llvm-project/}. Note in case you decide to manually build --  be prepared for LLVM to take up 130GB of disk and requires 8+ CPUS and 16+ GB of RAM. 

We have also already registered Wasm as a miscellaneous binary format as specified in the README.md --- you can just run \texttt{.wasm} files like ELF files (e.g. \texttt{./bash.wasm --norc})

\paragraph{wali-eurosys25-data}
The \texttt{README.md} in the \sysname experiment repo has a detailed setup guide as well. 
The only step required for setup is to move the needed \texttt{virt.tar.gz} binaries under \texttt{artifacts} with \textit{tar -xvf artifacts/virt.tar.gz -C wali-eurosys25-data/} for experiments.

\subsection{Evaluation workflow}
Below we highlight list the claims and required experiments needed to verify them.

\subsubsection{Major Claims}

\begin{itemize}
    \item \textit{(C1): \sysname allows porting most complex Linux applications unmodified over Wasm, unlike WASI/X. This is shown by E1 described in Sec.~\ref{sec:porting_effort} and Table.~\ref{table:wali_compiled}
    }
    \item \textit{(C2): \sysname allows layering complex security policies like \texttt{WASI} over it. This is shown by E2 described in Sec.~\ref{sec:porting_effort} and Table.~\ref{table:wali_compiled}.}
    \item \textit{(C3): \sysname strikes a middle-ground between container virtualization (Docker) and ISA virtualization (QEMU), with a reasonably fast startup overhead like QEMU, and a reasonably fast execution overhead like Docker. This is shown by E3 described in Sec.~\ref{sec:extrinsic-costs} and Fig.~\ref{fig:overhead}.}
\end{itemize}

\subsubsection{Experiments}

\paragraph{Experiment (E1)} \textit{[Portability] [20 human-minutes + 5 compute minutes]:} 
All ported \sysname Wasm apps are in \textit{applications-artifact.tar.gz}, which can be run directly. No preparation is necessary, and the ecosystem is already setup.

\textit{[Execution]}:
You can run all \sysname binaries like ELF binaries! Command line arguments are passed akin to normal binaries, and additionally a \texttt{WALI\_VERBOSE} environment variable can be configured to show dynamically executed syscalls
(e.g. \texttt{WALI\_VERBOSE=5 ./bash.wasm --norc}).

\textit{[Results]}:
As stated in Sec.~\ref{sec:porting_effort} and Table.~\ref{table:wali_compiled}, we can observe the dynamic syscalls executed that are missing from WASI/X in verbose mode -- the latter's spec doesn't support these so they wouldn't compile in the first place.

\paragraph{Experiment (E2)} \textit{[WASI Layering] [5 human-minutes + 5 compute-minutes]:}
We compile a popular implementation of WASI, \texttt{libuvwasi} to run over \sysname completely, passing all unit tests. 

\textit{[Execution]}:
Run the following commands:
\begin{enumerate}
    \item \texttt{cd WALI/applications}
    \item \texttt{./run\_libuvwasi\_tests.sh}
\end{enumerate}

\textit{[Results]}: 
The \texttt{ctest} harness should execute 22 tests, passing all of them as stated in Sec.~\ref{sec:porting_effort}, hence realizing Fig.~\ref{fig_wali_layering}.

\paragraph{Experiment (E3)} \textit{[Comparison to Docker/QEMU] [15 human-minutes + 30 compute-minutes]:}
Compare \sysname against Docker for container virtualization and QEMU for ISA virtualization for memory and runtime overheads.

\textit{[Execution]}:
Follow the instructions under \textit{Rerun Benchmarks} in the README for \sysname Experiments Repo for the relevant comparison. After executing, run \texttt{./gen\_plots.sh} to generate all the figures

\textit{[Results]}:
Since these tests are very sensitive to hardware platforms, the exact ratios may differ substantially between platforms. 
However, it should be consistent that \sysname strikes the middle ground, with faster startup time and memory usage than Docker and faster execution time than QEMU (Sec.~\ref{sec:extrinsic-costs}) -- see figures \texttt{memory.pdf} and  \texttt{runtime\_*.pdf}, which are similar to Fig.~\ref{fig:overhead})

\begin{table}[ht!]
  \small
  \begin{tabularx}{\columnwidth}{|X|X|}
    \hline
    \textbf{Filename} & \textbf{Description} \\
    \hline
    \texttt{memory.pdf} & Memory overhead plot (Fig.~\ref{fig:overhead-memory}) \\
    \hline
    \texttt{runtime\_*.pdf} & Runtime overhead plots (Fig.~\ref{fig:overhead-runtime-lua},~\ref{fig:overhead-runtime-bash},~\ref{fig:overhead-runtime-sqlite}) \\
    \hline
    \texttt{syscall\_profile.pdf} & Syscall Profile across applications (Fig.~\ref{fig_app_syscall_profile}) \\
    \hline
    \texttt{wali\_macrobench.pdf} & Macrobenchmark overhead split between \sysname, kernel and application space (Fig.~\ref{fig_wali_macrobench_overhead}) \\
    \hline
    \texttt{syscall\_archs.pdf} & Common syscalls between architectures (Fig.~\ref{fig_wali_arch_syscalls}) \\
    \hline
    \texttt{sigpoll.txt} & Sigpoll overhead results (Table.~\ref{table_wali_signal_overhead}) \\
    \hline
    \texttt{syscall\_overheads.txt} & Intrinsic WALI overhead (Table.~\ref{table_wali_syscall_overhead}) \\
    \hline
  \end{tabularx}
  \caption{List of generated plots from \texttt{data.tar.gz}. Generate all of these by running \texttt{./gen\_plots.sh}}
  \label{table:artifact-experiment-figures}
\end{table}
\subsection{Notes on Reusability}
\label{sec:reuse}

Generating all of the figures based on collected data (i.e those in Table.~\ref{table:artifact-experiment-figures}) is easy, and detailed in the README in \sysname Experiments Repo. 
Setting up the entire \sysname toolchain locally is also straight-forward, following the README in \sysname Code Repo.
We hope this encourages people to experiment and further research into Wasm over \sysname.

\subsection{General Notes}
\label{sec:gnotes}
\sysname is fully open-source and actively under development and improvement at the date of writing.

%
\end{document}